\documentclass[twocolumn,preprintnumbers,amsmath,amssymb,showpacs]{revtex4}

\pdfoutput=1 

\usepackage{graphicx}
\usepackage{bm}
\usepackage{dcolumn}
\usepackage{amssymb}
\usepackage{amsmath}
\usepackage{amsfonts}
\usepackage{color}
\usepackage{epsfig}

\def \Z2{$\mathbb{Z}_2$}

\def \be{\begin{equation}}
\def \ee{\end{equation}}
\def \ba{\begin{array}}
\def \ea{\end{array}}
\def \bea{\begin{eqnarray}}
\def \eea{\end{eqnarray}}

\begin{document}

\title{
     The $Z_2-$anomaly and boundaries of topological insulators }
\author{
    Zohar Ringel and Ady Stern}
\affiliation{
     Department of Condensed Matter Physics, Weizmann Institute of Science, Rehovot 76100, Israel}

\begin{abstract}
We study the edge and surface theories of topological insulators from the perspective of anomalies and identify a novel $Z_2-$anomaly associated with charge conservation. The anomaly is manifested through a 2-point correlation function involving creation and annihilation operators on two decoupled boundaries. Although charge conservation on each boundary requires this quantity to vanish, we find that it diverges. A corollary result is that under an insertion of a flux quantum the ground state evolves to an exactly orthogonal state independent of the rate at which the flux is inserted. The anomaly persists in the presence of disorder and imposes sharp restrictions on possible low energy theories. Being formulated in a many-body, field theoretical language, the anomaly allows to test the robustness of topological insulators to interactions in a concise way.
\end{abstract}

\maketitle

Topological insulators have attracted much attention in recent years due to their novel
bulk and surface properties. The bulk of these materials is insulating and characterized by topological indices which measure certain twists in the band structure. The topological properties  of the bulk imply, via the bulk-edge correspondence, that the surfaces of these materials are necessarily metallic and support gapless excitations.
The theories that emerge on the $d-1$ dimensional surfaces of a $d$-dimensional topological insulator may be understood as "fractions" of the theories of "stand-alone" $d-1$ dimensional systems \cite{Qi2008,Liu2012}. For example, the edge of an Integer Quantum Hall (IQHE) at filling factor one, arguably the simplest and most striking topological state of matter, may be thought of as half a spinless one-dimensional wire.

The surface theories of topological phases have several important properties, which are clearly demonstrated for the case of the IQHE. Although IQHE edges border between two insulating gapped phases, they appear to violate a conservation law of the entire system, namely charge conservation. Indeed the low energy theory of each edge is invariant under the $U(1)-$charge symmetry of the bulk. However, the application of an electric field parallel to the edges of a Hall bar leads to current flow across the bar. Consequently charges are exchanged between the edges and charge conservation, per-edge, is violated. This property is known as an anomaly and will occupy an important part of our discussion. In this case it is sometimes referred to as the Schwinger Anomaly \cite{Witten1984} or a $1D-$Chiral anomaly \cite{DHLee1996}.

While technically anomalies appear as a subtle cut-off scale effect, they are in fact present at all energy scales and have quite direct consequences \cite{Nakahara}. For example the above Schwinger anomaly fixes the commutation relations between the density operators on the edge \cite{Stone1991} to a quantized non-zero value, while naively one may think that density operators at different points on the edge commute. Ignoring the anomaly in this case, one may arrive at the wrong impression that the edge conductance vanishes, as if the edge was insulating, instead of being $e^2/h$.

The identification of the anomaly associated with an edge or surface of a topological state is particularly easy for the case of IQHE, where charge conservation is being violated or for the spin Quantum Hall Effect (QSHE), where spin conservation is violated. Generalizations to other topological phases have been explored \cite{Ryu2012,Chung2012} however no anomaly was found to be associated with $Z_2$ topological insulators (TI) in $2D$ or $3D$. In this respect it is important to note an anomalous behavior, mathematically similar to the $SU(2)$ global gauge anomaly \cite{Witten1982}, associated with the $O(2N) / O(N)\times O(N)-$symmetry of the disorder-averaged action in the replica formalism \cite{Ryu2007}. However, since the symmetry is not a gauge symmetry, this anomalous behavior does not imply a symmetry-violation or an inconsistency in the theory. Thus, it remained unclear whether any symmetry becomes anomalous in $Z_2$ topological insulators.

In this work we study topological insulators from the perspective of anomalies and identify a novel $Z_2$-anomaly that is associated with charge conservation on the boundary. The anomaly unifies the various aspects of topological insulators in a concise way through a field theoretical language and allows us to calculate new topological properties of TIs. To formulate the problem we consider the quantum action of two decoupled boundary theories, corresponding to two distinct edges or surfaces of a TI, and consider the cut-off scale as a bulk gap. We then dynamically insert a flux quantum so that an electric field along the boundaries is generated. Although the two boundary theories remain formally decoupled, we find that certain quantities are exchanged via the anomaly. This is manifested in two ways. First, regardless of how adiabatically the flux insertion is carried out, the final state is always an excited states which is exactly orthogonal to the ground state. Second, even without any direct coupling between the edges, certain 2-point correlation functions involving creation and annihilation operators on different edges diverge rather than vanish.

We begin by reformulating some well known results of the IQHE, at filling factor $1$, in a way that allows a generalization to TIs. Consider such an IQHE bar which is periodic in the $x-$direction and finite in the $y-$direction, i.e., an annulus. The topological nature of the bulk is manifested by two chiral edge states with opposite velocity which appear on the two disconnected edges. The basic Hamiltonian describing each edge is simply
\begin{align}
{\rm H}_{edge} &= v_0 (i\hbar \partial_x - e A_x)+\mu,
\end{align}
where $\mu$ is the chemical potential and we allowed coupling to a gauge field ($A_x$). Below we study in detail the case where the edges are identical expect having opposite velocities. Generalizations follow readily from the topological nature of our results.

The spectrum of ${\rm H}_{edge}$ performs a spectral-flow as a function of flux $(A_x)$ so that the states at $A_x = 0$ are exactly those at $A_x = h / (e L)$ displaced by a single state, where $L$ is the edge circumference. Thus as one inserts a flux quantum adiabatically, exactly one electron is transferred between the edges even though ${\rm H}_{edge}$, on each edge, has a $U(1)$-charge symmetry.  The spectral flow therefore implies the aforementioned charge anomaly of the IQHE edge.

We find it convenient to introduce a formalism that treats both edges simultaneously, and express this anomaly as a $2D-$chiral anomaly. To this end consider the action of {\bf both} the edges following a multiplication of $\bar{\psi}$ by $i\sigma_x$
\begin{align}
S &= \int dx d\tau \bar{\psi}_{\sigma} [\hat{{\rm S}}_{ch}]_{\sigma,\sigma'} \psi_{\sigma'}, \\
\label{Eq:SChiral}
\hat{{\rm S}}_{ch} &= (\alpha i\hbar \partial_{\tau} + i\mu)\sigma_x  + v_0 \sigma_y (i\hbar \partial_x -e A_x) \equiv \left( \begin{array}{cc}
0 & D \\
D^{\dagger} & 0 \\
\end{array} \right)
\end{align}
where the $\sigma-$spinor is associated with the edge index, $\bar{\psi},\psi$ are Grassman variables and $\alpha=1(i)$ for Euclidean (real-time) action. For simplicity we choose $A_x$ to be independent of $x$. Furthermore it couples symmetrically to the edges since we do not allow time-dependent magnetic fields in the bulk. The action operator, $\hat{{\rm S}}_{ch}$, acts in an extended Hilbert space which includes the extra edge index and the time coordinate. Notably the $U(1)$ symmetry associated with charge difference between the edges is now reflected by the fact that $\{\hat{{\rm S}}_{ch},\sigma_z\} = 0$, which we from now on refer to as the chiral symmetry.

Chiral symmetries are often anomalous \cite{Witten1984} and the former is no exception. Using Noether's procedure one finds that the chiral current is
\begin{align}
\vec{j}_{ch} &= e v_0 \bar{\psi} \sigma_z \vec{\sigma} \psi,
\label{chiral_current}
\end{align}
where $\vec{\sigma}=(\sigma_x,\sigma_y)$. A naive application of Noether's procedure will imply that the divergence of this current in space-time $\langle \partial_\mu {j}_{ch,\mu} \rangle$, with $\mu=x,\tau$, vanishes. This however is not necessarily the case in the presence of gauge-fields due to changes in the path integral measure \cite{Nakahara,Fujikawa1979}. Instead one finds the following fundamental relation of the chiral anomaly \cite{Nakahara,Stone1984},
\begin{align}
\label{Eq:Z-Anomaly}
\int d\tau dx \langle \vec{\nabla} \vec{j}_{ch}  \rangle  &= Spectral~flow = \frac{e}{h} \int d\tau dx F \\ \nonumber
\end{align}
where $F_{\mu \nu} = \partial_{t} A_x$, $A_x$ is periodic, up to gauge transformations, between $-\infty$ and $\infty$ (real-time) or $-\beta/2$ to $\beta/2$ (Euclidean-time). By $Spectral~flow$ we denote the integer number characterizing the displacement of the spectrum of ${\rm H}_{edge}[A_x(\tau)]$ between $\tau=-\infty$ and $\tau=\infty$. For example, $Spectral~flow = 1(-1)$ implies that when following an eigenvalue of ${\rm H}_{edge}[A_x(\tau=-\infty)]$ up to $\tau = \infty$, one ends up with the upper (lower) consecutive eigenvalue of the initial one.

Equation (\ref{Eq:Z-Anomaly}) clearly reflects the physics of the IQHE: The charge transferred between the edges is related, in a quantized fashion, to the flux inserted into the annulus. This relation is easy to generalize for the spin quantum Hall effect, in which the chiral charge current (\ref{chiral_current}) is replaced by a chiral spin current. However,  for a generic $2D$ TI there does not seem to be any local conservation law which is violated by an anomaly \cite{Ryu2012}.

Chiral anomalies are also accompanied by the appearance of action zero-modes which do generalize to TIs and have important physical consequences. The zero modes of the current model can be understood by viewing $\hat{\rm S}_{ch}$ as a Dirac Hamiltonian on a $2D$ torus (spanned by $(x,\tau)$) with a uniform magnetic field. In this system a zero energy Landau level appears with a degeneracy equal to the number of magnetic monopoles \cite{Jackiw1984}. This relation turns out to generalize to any chiral operator coupled to a gauge field via the Atiyah-Singer (AS) and Atiyah-Patodi-Singer (APS) theorems \cite{Nakahara}
\begin{align}
\label{Eq:AS}
\nu &=  Spectral~flow= \frac{e}{h}\int d\tau dx F,
\end{align}
where the analytic index $\nu$ is given by $\nu = DimKer[D] - DimKer[D^{\dagger}]$ and $DimKer[D]$ ($DimKer[D^{\dagger}]$) is the dimension of the zero mode space of $D$ ($D^{\dagger}$). Note that a zero mode of $D$ ($D^{\dagger}$) is also a zero mode of the action ($\hat{S}_{ch}$) with $\sigma_z = 1(-1)$ as can be verify explicitly from Eq. (\ref{Eq:SChiral}). Generalizations of all of our results to the $\mu\neq 0$ case (non-hermitian action), are given in App. (\ref{App:ZeroModes}).

Let us now generalize Eq. (\ref{Eq:AS}) to TIs. Consider a $2D$ TI cylinder with two distinct edges and again write the action for the two edges after performing the chirality transformation ($\bar{\psi} \rightarrow i\sigma_x \bar{\psi}$). This leads to a chiral action of the form
\begin{align}
\hat{{\cal S}}_{ch} &= (\alpha i\hbar \partial_{\tau} + i\mu)\sigma_x + \sigma_y {\cal H}_{edge}[A_x] = \left( \begin{array}{cc}
0 & {\rm D} \\
{\rm D}^{\dagger} & 0 \\
\end{array} \right),
\end{align}
where ${\cal H}_{edge}$ denotes the low energy Hamiltonian of a single TI edge, which we keep general. Provided that $A_x(\tau) = -A_x(-\tau)$, the action is invariant under time reversal symmetry (TRS)
\begin{align}
\label{Eq:TRSop}
T \hat{{\cal S}}_{ch} T^{-1} &=(is_y \sigma_x P_{\tau}) [\hat{{\cal S}}_{ch}]^T (i s_y \sigma_x P_{\tau})^T
\end{align}
where $s_y=\left (\begin{array}{cc}0 & -i\\  i & 0\end{array}\right )$ is the Pauli matrix acting on the electron spin and $P_{\tau}$ flips the sign of $\tau$. Since $T^2 = -1$ Kramer's theorem holds and each action eigenvalue is doubly degenerate. This again remains true also for the non-hermitian case in the sense that each Jordan block has an even dimension (see App. (\ref{App:ZeroModes})).

The presence of Kramer's theorem for the action allows us to define a $Z_2$ analytical index similar to $\nu$. To see this let us first assume ${\cal H}_{edge}$ respects $s_z-$symmetry and decouple the system into spin-up and spin-down components. When there is a single pair of zero modes, the two modes will have opposite $s_z$ and $\sigma_z$ (see definition of $T$). If we now break this spin symmetry, the zero modes and every other mode must remain degenerate due to TRS. The combination of the pairwise degeneracy of all modes along with the anti-symmetry of the spectrum around zero energy (implied by the chiral symmetry) allows the degeneracy of zero modes to change only in groups of four. This provides for the following topological index
\begin{align}
\label{Eq:nu2def}
\nu_2 &= DimKer[{\rm D}]~ mod ~ 2 = DimKer[{\rm D}^{\dagger}] ~ mod ~ 2,
\end{align}

The TIs have an analog to the spectral flow of the IQHE edges in the form of "pairswitching" \cite{Fu2006}. The "pair" refers here to the Kramer's pairs of edge states that characterize the spectrum at values of the flux that are time reversal symmetric, namely zero and one-half flux quanta. The term "switching" refers to the fact that states change their Kramer partners between zero flux and half a flux quantum. As we show below, this pairswitching behavior is related to the $\nu_2-$index of the action via
\begin{align}
\label{Eq:APSZ2}
\nu_2 &= Pairswitching
\end{align}
The r.h.s. is equal to $1$ ($0$) if the spectrum of ${\cal H}_{edge}[A_x]$ performs (does not perform) pairswitching as a function of $A_x(\tau)$. This generalization of the APS theorem relates the spectral motion characteristic of TIs boundaries with action zero modes in the presence electric fields.

To prove the relation in Eq. (\ref{Eq:APSZ2}), let us study the evolution of the action zero modes with $\tau$ for $A_x(\tau) = \frac{h\tau}{e L \beta}$. If a zero mode, $[\varphi_0(r,\tau),0]^T$, with $\sigma_z = +1$ exists, the following equation must be satisfied
\begin{align}
i {\rm D} \varphi_0 &= \left[\alpha \partial_{\tau} + {\cal H}_{edge}[A_x(\tau)]-\mu\right] \varphi_0 = 0.
\label{zeromodeeq}
\end{align}
with the boundary conditions,
\begin{align}
\varphi_0(r,\tau=-\beta/2) &= -\varphi_0(r,\tau=\beta/2)e^{-2 \pi i x/L}.
\label{zeromodebc}
\end{align}

To be concrete say we work with $\alpha=1$ (Euclidean action) and take an adiabatic limit so that $\beta \rightarrow \infty$ (the topological protection of the zero modes will later allow us to change these parameters). Provided that $\beta$ is large enough the flux insertion can viewed as the adiabatic evolution of $\varphi_0$ in imaginary time. While it is not obvious that one may still use the adiabatic theorem in imaginary time, this turns out to be true under some limitations \cite{Hwang1977}. The (imaginary) adiabatic theorem then implies
\begin{align}
\label{Eq:Adiabatic}
\varphi_0(x,\tau) &= \sum_n c_n e^{-\int d\tau (E_n(\tau)-\mu)} \psi_{n}(x;\tau), \\ \nonumber
{\cal H}_{edge}[A_x(\tau)]&\psi_{n}(x;\tau) = [E_n(\tau)-\mu] \psi_n(x;\tau).
\end{align}

To relate Eq. (\ref{Eq:Adiabatic}) with Eq. (\ref{Eq:APSZ2}) two observations are important. First, pairswitching and TRS imply that there are an odd number of time points, denoted by $\tau_{n}$, for which a state $E_m$ exists such that $E_{m}(\tau_{n})=\mu$ and $\partial_{\tau} E_{m}(\tau_{n})$ is positive. Second, upon choosing $c_n = \delta_{nm}$, for each such point the adiabatic evolution yields
\begin{align}
\varphi_0(r,\tau) &= e^{-\partial_{\tau} E_{m}(\tau_{n}) (\tau-\tau_{n})^2} \psi_{m}(r;\tau) + O(e^{-\beta \Delta}), \\ \nonumber
\end{align}
where $\Delta$ is the level spacing between two consecutive states of ${\cal H}_{edge}$.
In the adiabatic limit such states are localized near $\tau_{n}$ and trivially satisfy the boundary conditions in Eq. (\ref{zeromodebc}). Therefore an odd number of solutions to Eq. (\ref{zeromodeeq}) exist.
Looking for zero-modes with $\sigma_z = -1$, one obtains a variant of Eq. (\ref{zeromodeeq}) with the sign of ${\cal H}_{edge}$ switched. Repeating the above analysis yields another odd set of zero modes. According to Eq. (\ref{Eq:nu2def}) this implies $\nu_2 = 1$

In contrast, without pairswitching the edge spectrum ($E_m(\tau)$) crosses $\mu$, with a positive slope, an even number of times. Consequently, in the adiabatic limit, the $e^{-\int d\tau (E_m(\tau)-\mu)}-$factor in Eq. (\ref{Eq:Adiabatic}) guarantees an even number of zero modes for either ${\rm D}$ or ${\rm D}^{\dagger}$ yielding $\nu_2=0$ according to Eq. (\ref{Eq:nu2def}).

Having established the stability of zero-modes and their relation to pairswitching, we wish to point at their physical consequence. To this end consider a TI in its ground state on an annulus, in which the flux threading the hole is adiabatically varied from $-\Phi_0/2$ to $\Phi_0/2$ ($\Phi_0 = h/e$). We will now show that as long as the rate at which the flux is turned on is much smaller than the cut-off, the TI will end up being in a state that is exactly orthogonal to the ground state. Let us denote the ground state with $-\Phi_0/2$ by $|gs\rangle$ and consequently the ground state at $\Phi_0/2$ is $G|gs\rangle$, where $G=\exp^{i\frac{2\pi}{L}\int dx x\psi^+(x)\psi(x)}$. The state to which the system evolves after the flux is inserted is $U|gs\rangle$, with $U$ being the time evolution operator. We examine the overlap $\langle gs|G^+U|gs\rangle$ which, as shown in App. (\ref{App:Overlap}), can be written as the following path integral
\begin{align}
\langle gs| & G^{\dagger}U | gs \rangle = Z, \\ \nonumber
Z &= \int {\rm D}[\bar{\psi}\psi] e^{ - S[\bar{\psi},\psi]}, \\ \nonumber
S &= \int dtdx \bar{\psi} [\hat{{\cal S}}_{ch}(\alpha(t))] {\psi}+\int dt \alpha(t) E_{gs},
\end{align}
where $E_{gs}$ is the many-body ground state energy, the limit of $\beta \gg \Delta^{-1}$ is assumed and we take $\alpha(t)=i$ for the period of the flux insertion ($t \in [-\Delta t,\Delta t]$) and $\alpha=1$ for $t \in [-\beta/2,\Delta t],[\Delta t,\beta/2]$. As shown in App. (\ref{App:Overlap}), the boundary conditions are again those in Eq. (\ref{zeromodebc}).

The path-integral expression can be formally evaluated \cite{Pauli-Villars} using the action eigenvalues to yield
\begin{align}
\label{Eq:OverlapToZero}
\langle gs| G^{\dagger}U | gs \rangle &= Det[\hat{{\cal S}}_{ch}] = \Pi_n \beta \lambda_n,
\end{align}
where $\lambda_n$ are the eigenvalues (or the Jordan eigenvalues) of $\hat{{\cal S}}_{ch}$. Clearly in the presence of action zero modes this overlap vanishes. In the absence of zero modes and provided that the flux insertion rate is sufficiently small, the adiabatic theorem implies that this overlap is one, up to phases. Thus the overlap serves as a sharp distinction between a topological and trivial insulator, formulated in a many body language.

For the QSHE, the vanishing overlap follows from the change in spin on each of the edges, following the insertion of flux. For a generic TI, this behavior is much less obvious. Nonetheless in an extreme adiabatic limit, in which the flux insertion rate is smaller than $\Delta$, it may be understood by elementary means. Consider the spectrum of a TI edge and how it evolves with flux as it increases from $-\Phi_0/2$ to $\Phi_0/2$. At $\Phi_0/2$, all the states below the chemical potential are occupied. Following the pairswitching motion with flux, one finds that the ground state is degenerate at zero flux. This crossing of many-body states is not avoided due to TRS. Therefore, increasing the flux to $\Phi_0/2$ yields an excited state, orthogonal to the original ground state.

Last let us examine the effect zero modes have on the chiral symmetry which, at least naively, implies the conservation of charge-difference between the edges. Consider the inter-edge two-point correlation function, $G_{hop}= \langle \int dx dt  \bar{\psi}\sigma_0 s_0 \psi \rangle$,  where the average is taken with the chiral action. Since $G_{hop}$ switches sign following a $\pi-$chiral rotation, it should seemingly vanish. However, a direct calculation yields a different result. To show this, we include a source term $\hat{{\cal S}}_{ch}(m) = \hat{{\cal S}}_{ch} - m \bar{\psi}\sigma_0 s_0 \psi$, which physically amounts to coupling the charges on both edges, and find that
\begin{align}
\label{Eq:ExpectationValue}
&G_{hop}(m) \equiv \partial_m log\left[\int {\rm D}[\bar{\psi}\psi] e^{ -\int dt dx \bar{\psi}\hat{{\cal S}}_{ch}(m)\psi}\right] \\ \nonumber
&= \partial_m log\left(\Pi_n \beta [\lambda_n + m]\right) = \frac{2 \nu_2}{m} + O(m^0,m^1, ...) \\ \nonumber
\end{align}
Interestingly, in the presence of zero modes, this two-point function diverges as the edges become more and more decoupled, rather than vanishes. For this reason we say that the chiral symmetry has an anomaly. Furthermore the appearance of $2 \nu_2$ as the coefficient of $1/m-$pole, shows that this is indeed a $Z_2$ anomaly (see Eq. (\ref{Eq:nu2def})) so that coupling two anomalous systems yields a non-anomalous system.

In quantum mechanical language, the anomalous correlation function can be expressed as a diverging weak value \cite{Aharonov2005}. To this end we note that following standard relations between coherent state path integrals and operators formalism one finds
\begin{align}
G_{hop} &= \int_{-\Delta t}^{\Delta t} dt \frac{\langle Ggs(t)|[\psi_{top}^{\dagger} \psi_{bottom} + h.c]|gs(t)\rangle}{ \langle Ggs(t)|gs(t) \rangle},
\end{align}
where $|gs(t)\rangle=U^t_{-\Delta t}|gs \rangle$, $|Ggs(t) \rangle = [U^{\Delta t}_{t}]^{-1} G|gs\rangle$, $U_{a}^{b}$ is the unitary evolution between $t=a$ and $t=b$. Thus $G_{hop}$ appears as the time-averaged weak measurement of the inter-edge charge-tunneling operator. Physically this means that if one measures the charge-tunneling operator using a weakly coupled detector while selecting only the events in which the ground state came back to itself after the flux insertion- one will obtain a diverging value rather than a zero value.

So far we have addressed the case of the $2D$ TI however generalization of this anomaly to weak and strong TIs readily follow. Weak TIs can be understood as layers of $2D$ TIs and consequently, following its $Z_2-$nature, the anomaly will either be absent or present depending on the number of layers. For strong TIs, pairswitching on the surface depends on two fluxes. Namely one flux induces pairswitching conditioned on the value of the other flux \cite{FuKaneMele}. Consequently the anomaly, which follows directly from the pairswitching behavior, will be present or absent depending on whether this other flux is $0$ or $\Phi_0/2$.

The $Z_2$ charge anomaly identified in this work, may facilitate our understanding of topological insulator in the presence of disorder and interactions. The anomaly, which persists also in the disordered case, guarantees a non-localized phase on the boundary of TIs. This is because anomalous theories cannot be massive or have only short range correlations at low energy \cite{Frishman1981}. The anomaly should also be present in the Non-linear-sigma-model (NLSM) descriptions of the disordered surfaces \cite{Mirlin2007,Ryu2007,FuKane2012} and may, through the anomaly matching concept, allow further investigation of these theories. Considering interactions, it remains unclear whether topological insulators remain well defined. It is therefore interesting to see whether Eq. (\ref{Eq:ExpectationValue}), which sharply distinguish a topological surface from a trivial one, receives any perturbative interaction corrections.

We thank the US-Israel Binational Science Foundation and the Minerva foundation for financial support.

\appendix

\section{The anomaly in the non-Hermitian case}
\label{App:ZeroModes}
The chiral euclidean-action is an hermitian operator only for $\mu=0$ and more generally, only when edge symmetric terms are absent. Here we show that the results derived in the main text persist also when such terms are included. We first define the analytic index ($\nu$) for a non-hermitian action and prove its stability. Next a variant of Kramer's theorem is proven and used for showing the robustness of the $Z_2$-analytic index ($\nu_2$).

A general chiral action is given by
\begin{align}
S &= \left(\begin{array}{cc}
0 & D_- \\
D_+ & 0 \\
\end{array} \right),
\end{align}
where $D_+$ and $D_-$ are not necessarily hermitian conjugates. The action, when viewed as a matrix, might be non-diagonalizable so that no spanning basis of eigenvectors exists, not even a non-orthogonal one. Nonetheless a weaker statement holds which is that any matrix can be brought to a Jordan form ($\tilde{S}$) following a similarity transformation ($\tilde{S} = P S P^{-1}$). In its Jordan form \cite{Hoffman1971}, the action becomes block diagonal so that each block ($J_n$) has a single Jordan eigenvalue ($\lambda_n$) on its diagonal and either $1$ or $0$ on the upper off-diagonal. All other entries of $\tilde{S}$ are zero. Alternatively stated, a set of Jordan vectors exists ($v_{n,i}$)  upon which the matrix $(S-I\lambda_n)$ is nilpotent so that $(A-I\lambda_n)^{N_n} v_{n,i} = 0$, where $N_n \leq {\rm Dim}[J_n]$.

We extend the definition of $\nu$ to the non-hermitian as followed
\begin{align}
\nu &={\rm Dim}\left[J_0(D_- D_+)\right] - {\rm Dim}\left[J_0(D_+D_-)\right],
\end{align}
where $J_0(A)$ denotes the Jordan block associated with a zero Jordan eigenvalue of the matrix $A$. Let us show that for the hermitian case, the above definition is equivalent to $\nu =  DimKer[D^{\dagger}]-DimKer[D]$, used in the main text. Note that for an hermitian operator ${\rm Dim}\left[J_0(D D^{\dagger})\right]=DimKer[DD^{\dagger}]$ and similarly for the hermitian conjugate operators. Next note that if a vector, $v$, is in $Ker[D]$ ($Ker[D^{\dagger}]$) then it must also be in $Ker[D^{\dagger}D]$ ($Ker[D D^{\dagger}]$). Furthermore, if $v$ is in the kernel of $Ker[D^{\dagger}D]$ ($Ker[D D^{\dagger}]$) it must also be in $Ker[D]$ ($Ker[D^{\dagger}]$) since $D^{\dagger}Dv=0$ implies $v^{\dagger}D^{\dagger}Dv=|Dv|^2=0$ and therefore $Dv=0$. Consequently $Ker[D] = Ker[D^{\dagger}D]$ and $Ker[D^{\dagger}]=Ker[DD^{\dagger}]$ and the two definitions of $\nu$, for the hermitian case, are clearly equivalent.

In the hermitian case each vector in $Ker[D]$ ($Ker[D^{\dagger}]$) can also be used to build a zero mode of $S$ with $\sigma_z = 1$ ($\sigma_z = -1$) and thus the second definition of $\nu$ relates better to the zero modes of $S$. For the non-hermitian case, the following weaker relation holds between $\nu$ and zero modes of $S$
\begin{align}
{\rm Dim}\left[J_0(S)\right] \ge |\nu|.
\end{align}
To prove this last statement note that
\begin{align}
S^2 &= \left(\begin{array}{cc}
D_-D_+ & 0 \\
0 & D_+ D_- \\
\end{array} \right),
\end{align}
and therefore ${\rm Dim}\left[J_0(S^2)\right] \ge |\nu|$ and using Jordan form we have that ${\rm Dim}\left[J_0(S^2)\right] = {\rm Dim}\left[J_0(S)\right]$.

Next we show that $\nu$, as defined above, is a stable number. Let $v_{n,i}$ and $\lambda_n$ denote the Jordan vectors and eigenvalues of $D_+ D_-$. Acting with $D_-$ on $v_{n,i}$ gives us Jordan vectors of $D_-D_+$, as one can directly show. Furthermore, $D_-$ generates a one to one map between the Jordan vectors of these two operators for which $\lambda_n \neq 0$. Indeed, if there exists some linear combination $v_{n}$, of $v_{n,i}$ such that $D_- v_n = 0$, then also $D_+ D_- v_n = 0$ implying $\lambda_n=0$. Consider now a small perturbation which increases ${\rm Dim}\left[J_0(D_+ D_-)\right]$. This implies that a Jordan block with some small $\lambda_n$ appears for $D_+D_-$ and must be matched, through the mapping, to an equal size Jordan block appearing in $D_- D_+$. This later block can only appear from an equal change of ${\rm Dim}\left[J_0(D_- D_+)\right]$ and consequently $\nu$ is stable to any perturbation.

Next we wish to show that the following variant of Kramer's theorem holds also in the non-hermitian case. Consider an action which obeys a Fermionic TRS symmetry
\begin{align}
O^T S O &= S^T, \\
O &= -O^T, \\
O O^T &= 1,
\end{align}
where $O$ in the usual case is $is_y$. Then for such a matrix, $S$, each Jordan block has an even dimension.
To prove this let us show how this symmetry acts on the Jordan form. Note that
\begin{align}
S^T = [P^{-1} \tilde{S}P]^T &= P^T \tilde{S}^T [P^T]^{-1} = O^T S O,
\end{align}
and by acting with $[P^{-1}]^T$ and $P^T$ on the two sides of this last equality one finds
\begin{align}
\tilde{S}^T = [P^T]^{-1} O^T S O P^T &= [P^T]^{-1} O^T P^{-1} \tilde{S} P O P^T.
\end{align}
Denoting $\tilde{O} = P O P^T$ one has that
\begin{align}
\label{Eq:OonJordan}
 \tilde{S}^T &=  \tilde{O}^{-1} \tilde{S} \tilde{O}, \\
\tilde{O}^T &= -\tilde{O}.
\end{align}

Next we show that $\tilde{O}$ is block diagonal on the basis, $\hat{e}_{n,i}$, on which $\tilde{S}$ and $\tilde{S}^T$ are block diagonal. For each $\lambda_n$ block of size $N$ of $\hat{S}$, one can use Eq. (\ref{Eq:OonJordan}) to show that
\begin{align}
\tilde{O} (\tilde{S}^T-\lambda_n)^N = (\tilde{S} - \lambda_n)^N \tilde{O}.
\end{align}
Acting with $\hat{e}^T_{n,i}$ from the left and using $\hat{e}_{n,i}^T (\tilde{S}-\lambda_n)^N =0$ one finds
\begin{align}
\hat{e}^T_{n,i}\tilde{O} (\tilde{S}^T-\lambda_n)^N w &= 0,
\end{align}
for any vector $w$. Using the fact that $\tilde{S}^T - \lambda_n$ is invertible when constrained to the subspace of $\hat{e}_{m,j}$ with $m \neq n$, one may take $w = [(\tilde{S}^T-\lambda_n)^N]^{-1} \hat{e}_{m \neq n,j}$ and obtain
\begin{align}
\hat{e}_{n,i}^T \tilde{O} \hat{e}_{m,j} &= 0 \,\,\,\ \forall m \neq n,
\end{align}
showing that $\tilde{O}$ is indeed block diagonal on this basis.
Last using the fact that $\tilde{O}$ is invertible and anti-symmetric we find that
\begin{align}
\det[\tilde{O}_n] \neq 0 \Leftrightarrow {\rm Dim}[\tilde{O}_n] \in N_{even},
\end{align}
where $\tilde{O}_n$ is the $n$ block of $\tilde{O}$. Therefore each block of the Jordan matrix must have an even number of states and a variant Kramer's theorem holds even in the non-hermitian case.

The robustness of the $\nu_2$-index, for the non-hermitian case, follows readily from the definition
\begin{align}
\nu_2 &= ({\rm Dim}\left[J_0(S)\right] \,\,\ mod \,\,\ 4)/2.
\end{align}
Indeed using the chiral and TRS symmetry of the action one can show that this quantity can only change by multiples of $4$.

\section{Overlap as a path integral}
\label{App:Overlap}
Here we show that the partition function discussed in the main text is equal to an overlap of two distinct states: The ground state in the presence of a full flux quantum and the ground state to which we dynamically insert a flux quantum at a rate which is smaller than the cut-off scale or equivalently the bulk-gap.

To establish this result we rewrite the overlap, up to phases, as
\begin{align}
\label{AppEq:Exp2Trace}
\langle gs| G^{\dagger}U | gs \rangle &= Tr \left[P G^{\dagger}U P \right] = Tr \left[ G^{\dagger} [G P G ^{\dagger}] U \right]\\
P &= \lim_{\beta \rightarrow \infty} e^{-\beta (H[0]-E_{gs})}, \\
U &= e^{-i \int^{\Delta t}_{-\Delta t} (H[A(t)]-E_{gs})}, \\
G &=\exp^{i\frac{2\pi}{L}\int dx x\psi^{\dagger}(x)\psi(x)},
\end{align}
where $H[A]$ is the full system Hamiltonian in second quantization coupled to a gauge field $A$, $\Delta t$ is the duration of the flux insertion process, $E_{gs}$ is a c-number equal to the ground state energy and therefore $P$ projects on the ground state. The gauge field is given by $A(t)=\frac{e}{h}\frac{t}{2L \Delta t}$ for $t \in [-\Delta t,\Delta t]$ corresponding to the insertion of a single flux quantum.

Taking advantage of the fact that $P$ is presented as a time evolution operator we may unite $G P G^{\dagger},U$ and $P$ into a single time evolution operator
\begin{align}
G P G^{\dagger} U P &= e^{- \int^{\beta/2}_{-\beta/2}dt \alpha^{-1}(t) H[A(t)]}
\end{align}
where for $t \in [-\beta/2,-\Delta t]$ we take $A=-e/(2hL)$ and $\alpha^{-1}(t) = 1$. For $t \in [\Delta t,\beta/2]$ we take $A=+e/(2hL)$ and $\alpha^{-1}(t) = 1$. In the remaining interval, $t \in [-\Delta t,\Delta t]$, we take $A(t) = \frac{e}{h}\frac{t}{2 L \Delta t},\alpha^{-1}(t)=i$.

Using Grassmann calculus to take the the trace in Eq. (\ref{AppEq:Exp2Trace}) we obtain
\begin{align}
\langle gs| G^{\dagger}U | gs \rangle &= \int {\rm D}[\bar{c} c] e^{-\bar{c}c} \langle -c |G^{\dagger} e^{- \int^{\beta/2}_{-\beta/2} dt \alpha^{-1}(t) H[A(t)]} |c \rangle, \\
|c\rangle &= e^{-c_i a^{\dagger}_i} |0\rangle,
\end{align}
where $|0\rangle$ is the vacuum, $a_i^{\dagger}$ is a creation operator for some spanning basis of states indexed by $i$, $\bar{c}c = \sum_i \bar{c}_i c_i$, the limit $\beta \rightarrow \infty$ is assumed and summation over repeated indices is implicit.

Next we follow the usual steps of constructing a path integral from Fermionic coherent states (see for example Ref.~\cite{Altland}). First the time evolution is partitioned into a product of $N\rightarrow \infty$ infinitesimal time evolutions, next Grassman resolutions of the identity in the form $\int {\rm D}[\bar{\psi}_n \psi_n] e^{-\bar{\psi}_n\psi_n}|\psi_n\rangle \langle \psi_n|$ are inserted between each of those, last using $a_n |\psi_n\rangle = \psi_n | \psi_n \rangle$ the operator $H(a^{\dagger},a)$ is replaced by a Grassmanian functional.

The only non-standard complication in this path integral is the presence of the gauge transformation $G$. This turns out to be equivalent to a certain choice of boundary conditions as we now show. First we present the action of $G$ on a grassman coherent state $|c\rangle$ through its action on the grassman variables
\begin{align}
G | c \rangle &= e^{- c_i G a_i^{\dagger} G^{\dagger}} |0\rangle = e^{-c_i g_{ij} a_j^{\dagger}}|0\rangle, \\ \nonumber
G a_i^{\dagger} G^{\dagger} &= g_{ij}a^{\dagger}_j, \\ \nonumber
G a_i G^{\dagger} &= g^*_{ij}a_j,
\end{align}
so if $i$ is the position basis, $G a_i^{\dagger} G^{-1} = e^{i 2\pi x_i/L}a^{\dagger}_i$ and $g_{ij} = \delta_{ij} e^{i 2\pi x_i / L}$. Following this we obtain
\begin{align}
G|c\rangle &= |cg\rangle, \\ \nonumber
\langle cg | &= \langle c | G^{\dagger} = \langle 0| e^{-a_j \bar{c}_i g^{*}_{ij} }.
\end{align}

Boundary conditions for the path integral are determined by the requirement that the discrete time derivatives generated by the time slicing process remain finite. These time derivatives emerge from the $-\bar{\psi}_n \psi_n$ factors coming from the resolution of the identity and the $\bar{\psi}_n \psi_{n+1}$ coming from the overlap of adjacent coherent states. Focusing only on these elements, the path generates a series of the form
\begin{align}
e^{-\bar{c}c}&\langle -cg | \psi_0 \rangle  e^{ -\bar{\psi}_0\psi_0} \langle \psi_0 |\psi_1\rangle(...)e^{-\bar{\psi}_N\psi_N} \langle \psi_N | c \rangle \\ \nonumber
&= e^{-\bar{c}c - \bar{c} g^* \psi_0 - \bar{\psi}_0 \psi_0 + \bar{\psi}_0 \psi_1 + (...) - \bar{\psi}_N\psi_N + \bar{\psi}_N c}
\end{align}
This series can regrouped to form a sum of derivatives either by grouping adjacent elements that share $\bar{\psi}$ or elements that share $\psi$ (this ambiguity is a discrete version of integration by parts). Consequently we require both of the following expression to remain finite as we refine the time slicing
\begin{align}
&e^{-\bar{c}(c + g^* \psi_0) - \bar{\psi}_0 (\psi_0 - \psi_1) + (...) - \bar{\psi}_N(\psi_N - c)}  \\ \nonumber
&e^{(\bar{\psi}_N-\bar{c})c + (-\bar{c} g^* - \bar{\psi}_0) \psi_0 + (\bar{\psi}_0-\bar{\psi}_1) \psi_1 + (...)^{'} + (\bar{\psi}_{N-1}-\bar{\psi}_N)\psi_N}
\end{align}
Identifying $\psi(\beta/2)$ with $\psi_0$ and $\psi(-\beta/2)$ with $\psi_N$ the boundary conditions are
\begin{align}
\psi(-\beta/2) &= -g^* \psi(\beta/2)= -\psi(\beta/2)g^{\dagger}, \\
\bar{\psi}(-\beta/2) &= - \bar{\psi}(\beta/2) g^T = -g \bar{\psi}(\beta/2).
\end{align}
which in position basis amounts to
\begin{align}
\label{Eq:GrassmanBC}
\psi(-\beta/2,x) &= -e^{-i2\pi x/L}\psi(\beta/2,x),
\end{align}
and its complex conjugate condition.
The path integral expression for the overlap is therefore
\begin{align}
\langle gs| G^{\dagger}U | gs \rangle &= \int {\rm D}[\bar{\psi}\psi] e^{ -S[\bar{\psi},\psi]}, \\ \nonumber
S &=  \bar{\psi} \partial_t \psi - \alpha^{-1}(t) (H[\bar{\psi},\psi]-E_{gs}),
\end{align}
with field configurations obeying Eq. (\ref{Eq:GrassmanBC}). Taking $\bar{\psi} \rightarrow \bar{\psi} \alpha(t)$, yields
\begin{align}
S &=  \bar{\psi} \alpha (t) \partial_t \psi - H[\bar{\psi},\psi]+E_{gs},
\end{align}

In the main text we derived the zero modes associated with the Euclidean chiral action ($\alpha(t)=1$), however these also persist with the extra $\alpha(t)$-factor. Indeed one can start from $\alpha_0(t) = 1$ gradually deform it, while respecting TRS as defined in Eq. (\ref{Eq:TRSop}), to $\alpha(t)$. Such a continuous deformation cannot remove the topologically protected zero modes even though the operators associated with $S$ becomes non-Hermitian (see App. \ref{App:ZeroModes}).

\bibliography{Anomaly}

\begin{thebibliography}{23}
\expandafter\ifx\csname natexlab\endcsname\relax\def\natexlab#1{#1}\fi
\expandafter\ifx\csname bibnamefont\endcsname\relax
  \def\bibnamefont#1{#1}\fi
\expandafter\ifx\csname bibfnamefont\endcsname\relax
  \def\bibfnamefont#1{#1}\fi
\expandafter\ifx\csname citenamefont\endcsname\relax
  \def\citenamefont#1{#1}\fi
\expandafter\ifx\csname url\endcsname\relax
  \def\url#1{\texttt{#1}}\fi
\expandafter\ifx\csname urlprefix\endcsname\relax\def\urlprefix{URL }\fi
\providecommand{\bibinfo}[2]{#2}
\providecommand{\eprint}[2][]{\url{#2}}

\bibitem[{\citenamefont{Qi et~al.}(2008)\citenamefont{Qi, Hughes, and
  Zhang}}]{Qi2008}
\bibinfo{author}{\bibfnamefont{X.-L.} \bibnamefont{Qi}},
  \bibinfo{author}{\bibfnamefont{T.~L.} \bibnamefont{Hughes}},
  \bibnamefont{and} \bibinfo{author}{\bibfnamefont{S.-C.} \bibnamefont{Zhang}},
  \bibinfo{journal}{Phys. Rev. B} \textbf{\bibinfo{volume}{78}},
  \bibinfo{pages}{195424} (\bibinfo{year}{2008}).

\bibitem[{\citenamefont{{Liu} et~al.}(2012)\citenamefont{{Liu}, {Qi}, and
  {Zhang}}}]{Liu2012}
\bibinfo{author}{\bibfnamefont{C.-X.} \bibnamefont{{Liu}}},
  \bibinfo{author}{\bibfnamefont{X.-L.} \bibnamefont{{Qi}}}, \bibnamefont{and}
  \bibinfo{author}{\bibfnamefont{S.-C.} \bibnamefont{{Zhang}}},
  \bibinfo{journal}{Physica E Low-Dimensional Systems and Nanostructures}
  \textbf{\bibinfo{volume}{44}}, \bibinfo{pages}{906} (\bibinfo{year}{2012}),
  \eprint{1110.3420}.

\bibitem[{\citenamefont{Witten}(1984)}]{Witten1984}
\bibinfo{author}{\bibfnamefont{E.}~\bibnamefont{Witten}},
  \bibinfo{journal}{Communications in Mathematical Physics}
  \textbf{\bibinfo{volume}{92}}, \bibinfo{pages}{455} (\bibinfo{year}{1984}).

\bibitem[{\citenamefont{Kao and Lee}(1996)}]{DHLee1996}
\bibinfo{author}{\bibfnamefont{Y.-C.} \bibnamefont{Kao}} \bibnamefont{and}
  \bibinfo{author}{\bibfnamefont{D.-H.} \bibnamefont{Lee}},
  \bibinfo{journal}{Phys. Rev. B} \textbf{\bibinfo{volume}{54}},
  \bibinfo{pages}{16903} (\bibinfo{year}{1996}).

\bibitem[{\citenamefont{Nakahara}(2003)}]{Nakahara}
\bibinfo{author}{\bibfnamefont{M.}~\bibnamefont{Nakahara}},
  \emph{\bibinfo{title}{Geometry, Topology and Physics}}
  (\bibinfo{publisher}{IOP Publishing Ltd.}, \bibinfo{address}{Bristol and
  Philadelphia}, \bibinfo{year}{2003}).

\bibitem[{Sto(1991)}]{Stone1991}
\bibinfo{journal}{Annals of Physics} \textbf{\bibinfo{volume}{207}},
  \bibinfo{pages}{38 } (\bibinfo{year}{1991}), ISSN \bibinfo{issn}{0003-4916}.

\bibitem[{\citenamefont{{Ryu} et~al.}(2012)\citenamefont{{Ryu}, {Moore}, and
  {Ludwig}}}]{Ryu2012}
\bibinfo{author}{\bibfnamefont{S.}~\bibnamefont{{Ryu}}},
  \bibinfo{author}{\bibfnamefont{J.~E.} \bibnamefont{{Moore}}},
  \bibnamefont{and} \bibinfo{author}{\bibfnamefont{A.~W.~W.}
  \bibnamefont{{Ludwig}}}, \bibinfo{journal}{\prb}
  \textbf{\bibinfo{volume}{85}}, \bibinfo{eid}{045104} (\bibinfo{year}{2012}),
  \eprint{1010.0936}.

\bibitem[{\citenamefont{{Chung} et~al.}(2012)\citenamefont{{Chung}, {Horowitz},
  and {Qi}}}]{Chung2012}
\bibinfo{author}{\bibfnamefont{S.~B.} \bibnamefont{{Chung}}},
  \bibinfo{author}{\bibfnamefont{J.}~\bibnamefont{{Horowitz}}},
  \bibnamefont{and} \bibinfo{author}{\bibfnamefont{X.-L.} \bibnamefont{{Qi}}},
  \bibinfo{journal}{ArXiv e-prints}  (\bibinfo{year}{2012}),
  \eprint{1208.3928}.

\bibitem[{Wit(1982)}]{Witten1982}
\bibinfo{journal}{Physics Letters B} \textbf{\bibinfo{volume}{117}},
  \bibinfo{pages}{324 } (\bibinfo{year}{1982}), ISSN \bibinfo{issn}{0370-2693}.

\bibitem[{\citenamefont{{Ryu} et~al.}(2007)\citenamefont{{Ryu}, {Mudry},
  {Obuse}, and {Furusaki}}}]{Ryu2007}
\bibinfo{author}{\bibfnamefont{S.}~\bibnamefont{{Ryu}}},
  \bibinfo{author}{\bibfnamefont{C.}~\bibnamefont{{Mudry}}},
  \bibinfo{author}{\bibfnamefont{H.}~\bibnamefont{{Obuse}}}, \bibnamefont{and}
  \bibinfo{author}{\bibfnamefont{A.}~\bibnamefont{{Furusaki}}},
  \bibinfo{journal}{Physical Review Letters} \textbf{\bibinfo{volume}{99}},
  \bibinfo{eid}{116601} (\bibinfo{year}{2007}),
  \eprint{arXiv:cond-mat/0702529}.

\bibitem[{\citenamefont{Fujikawa}(1979)}]{Fujikawa1979}
\bibinfo{author}{\bibfnamefont{K.}~\bibnamefont{Fujikawa}},
  \bibinfo{journal}{Phys. Rev. Lett.} \textbf{\bibinfo{volume}{42}},
  \bibinfo{pages}{1195} (\bibinfo{year}{1979}).

\bibitem[{Sto(1984)}]{Stone1984}
\bibinfo{journal}{Annals of Physics} \textbf{\bibinfo{volume}{155}},
  \bibinfo{pages}{56 } (\bibinfo{year}{1984}), ISSN \bibinfo{issn}{0003-4916}.

\bibitem[{\citenamefont{Jackiw}(1984)}]{Jackiw1984}
\bibinfo{author}{\bibfnamefont{R.}~\bibnamefont{Jackiw}},
  \bibinfo{journal}{Phys. Rev. D} \textbf{\bibinfo{volume}{29}},
  \bibinfo{pages}{2375} (\bibinfo{year}{1984}).

\bibitem[{\citenamefont{Fu and Kane}(2006)}]{Fu2006}
\bibinfo{author}{\bibfnamefont{L.}~\bibnamefont{Fu}} \bibnamefont{and}
  \bibinfo{author}{\bibfnamefont{C.~L.} \bibnamefont{Kane}},
  \bibinfo{journal}{Phys. Rev. B} \textbf{\bibinfo{volume}{74}},
  \bibinfo{pages}{195312} (\bibinfo{year}{2006}).

\bibitem[{\citenamefont{Hwang and Pechukas}(1977)}]{Hwang1977}
\bibinfo{author}{\bibfnamefont{J.-T.} \bibnamefont{Hwang}} \bibnamefont{and}
  \bibinfo{author}{\bibfnamefont{P.}~\bibnamefont{Pechukas}},
  \bibinfo{journal}{J. Chem. Phys.} \textbf{\bibinfo{volume}{67}},
  \bibinfo{pages}{4640} (\bibinfo{year}{1977}).

\bibitem[{Pau()}]{Pauli-Villars}
\bibinfo{note}{The determinant can be made well defined using a Pauli-Villars
  regulator \cite{Fujikawa1979} which generates a $\Pi_n [\lambda_n-iM]$
  denominator.}

\bibitem[{\citenamefont{Aharonov and Botero}(2005)}]{Aharonov2005}
\bibinfo{author}{\bibfnamefont{Y.}~\bibnamefont{Aharonov}} \bibnamefont{and}
  \bibinfo{author}{\bibfnamefont{A.}~\bibnamefont{Botero}},
  \bibinfo{journal}{Phys. Rev. A} \textbf{\bibinfo{volume}{72}},
  \bibinfo{pages}{052111} (\bibinfo{year}{2005}).

\bibitem[{\citenamefont{Fu et~al.}(2007)\citenamefont{Fu, Kane, and
  Mele}}]{FuKaneMele}
\bibinfo{author}{\bibfnamefont{L.}~\bibnamefont{Fu}},
  \bibinfo{author}{\bibfnamefont{C.~L.} \bibnamefont{Kane}}, \bibnamefont{and}
  \bibinfo{author}{\bibfnamefont{E.~J.} \bibnamefont{Mele}},
  \bibinfo{journal}{Phys. Rev. Lett.} \textbf{\bibinfo{volume}{98}},
  \bibinfo{pages}{106803} (\bibinfo{year}{2007}).

\bibitem[{Fri(1981)}]{Frishman1981}
\bibinfo{journal}{Nuclear Physics B} \textbf{\bibinfo{volume}{177}},
  \bibinfo{pages}{157 } (\bibinfo{year}{1981}), ISSN \bibinfo{issn}{0550-3213}.

\bibitem[{\citenamefont{Ostrovsky et~al.}(2007)\citenamefont{Ostrovsky, Gornyi,
  and Mirlin}}]{Mirlin2007}
\bibinfo{author}{\bibfnamefont{P.~M.} \bibnamefont{Ostrovsky}},
  \bibinfo{author}{\bibfnamefont{I.~V.} \bibnamefont{Gornyi}},
  \bibnamefont{and} \bibinfo{author}{\bibfnamefont{A.~D.}
  \bibnamefont{Mirlin}}, \bibinfo{journal}{Phys. Rev. Lett.}
  \textbf{\bibinfo{volume}{98}}, \bibinfo{pages}{256801}
  (\bibinfo{year}{2007}).

\bibitem[{\citenamefont{{Fu} and {Kane}}(2012)}]{FuKane2012}
\bibinfo{author}{\bibfnamefont{L.}~\bibnamefont{{Fu}}} \bibnamefont{and}
  \bibinfo{author}{\bibfnamefont{C.~L.} \bibnamefont{{Kane}}},
  \bibinfo{journal}{ArXiv e-prints}  (\bibinfo{year}{2012}),
  \eprint{1208.3442}.

\bibitem[{\citenamefont{Hoffman and Kunze}(1971)}]{Hoffman1971}
\bibinfo{author}{\bibfnamefont{K.}~\bibnamefont{Hoffman}} \bibnamefont{and}
  \bibinfo{author}{\bibfnamefont{R.}~\bibnamefont{Kunze}},
  \emph{\bibinfo{title}{Linear Algebra}} (\bibinfo{publisher}{Prentice-Hall},
  \bibinfo{year}{1971}), ISBN \bibinfo{isbn}{9780135368213}.

\bibitem[{\citenamefont{Altland and Simons}(2010)}]{Altland}
\bibinfo{author}{\bibfnamefont{A.}~\bibnamefont{Altland}} \bibnamefont{and}
  \bibinfo{author}{\bibfnamefont{B.}~\bibnamefont{Simons}},
  \emph{\bibinfo{title}{Condensed Matter Field Theory}}
  (\bibinfo{publisher}{Cambridge University Press},
  \bibinfo{address}{Cambridge, UK}, \bibinfo{year}{2010}),
  \bibinfo{edition}{2nd} ed.

\end{thebibliography}

\end{document}